\documentstyle[12pt]{article}
\def\be{\begin{equation}}
\def\ee{\end{equation}}

\begin{document}
\title{
\vspace{2cm}
Models of Quantum Computers \\ and \\
Decoherence Problem}
\author{
I.V. Volovich \thanks{Invited talk at the International
Conference on Quantum Information, Meijo University, Nagoya,
4-8 Nov.1997.
To be published in Proc. of the Conference}
$~~~~$  \\
{\it Steklov Mathematical Institute, Russian Academy of Sciences}\\
{\it Gubkin St.8, GSP-1, 117966, Moscow, Russia}\\
volovich@mi.ras.ru}
\date {$~$}
\maketitle
\begin {abstract}

Ma\-the\-ma\-ti\-cal mo\-dels of quantum computers such as a
mul\-ti\-di\-men\-sio\-nal quan\-tum Tu\-ring
ma\-chi\-ne and quan\-tum circuits are described and its relations
with lattice spin models
are discussed. One of the main open problems one has to solve if one  wants
to build a  quantum computer is the decoherence
due to the coupling with the  environment.
We propose a possible solution of this problem by using a control
of parameters of the system.
This proposal is based on the analysis of the spin-boson Hamiltonian
performed in the stochastic limit approximation.
\end{abstract}

\newpage
\section{Introduction}

Quantum computation touches upon the foundations
of mathematics, quantum physics and computer science.
Various mathematical and physical aspects of quantum computations
have been discussed in many works, see for example \cite{Ben}-\cite{Oza}.
Feynman \cite{Fey} has pointed out that there is no efficient way
of simulating a quantum mechanical system on computer
and suggested that, perhaps, a computer based on
quantum principles might be able to carry out the simulation
efficiently. One hopes that remarkable properties of quantum systems such as
quantum parallelism and entanglement will lead to a major
breakthrough in computer science.

Quantum computers violate the modern form of the Church-Turing
thesis. It was shown that certain problems, in particular
the recursive Fourier sampling problem, can be solved in polynomial
time on a quantum Turing machine, but relative to an oracle,
requires superpolynomial time on a classical probabilistic
Turing machine \cite{BV}. An efficient quantum algoriphm for factorizing
integers was found by Shor \cite{Sho}.

 There exist already
small quantum computers working with few qubits. However
it is not yet clear whether large quantum computers
suitable for practical use are possible to build. One reason
that quantum computer is difficult to build is decoherence
of quantum superpositions due to the interaction with the environment.
Decoherence can be defined as the decay of the off-diagonal matrix
elements of the density operator in the computational basis.
 By using simple models like spin system coupling with quantum field
it was found that decoherence is quite large.
Different methods were discussed to reduce the decoherence
in quantum computers \cite{Unr}-\cite{DG}, still
it is considered to be as
one of the main open problems in this topic.

In this work we propose an approach to the problem of reducing
decoherence in quantum memory by using a control
on the parameters of the system \cite{Vol1}. We show that
one can choose the parameters of the system in such
a way that decoherence is drastically reduced.
This proposal is based on the previous analysis \cite{AKV}
of the spin-boson Hamiltonian
 performed in the stochastic limit approximation.
Considerations of the decoherence problem in quantum computers
performed in \cite{Unr,PSE,DG} used a special case of the
spin-boson interaction when no spin-flip transitions (or tunneling,
in another interpretation) are present. Here we consider the complete
spin-boson model, including the spin-flip term,
and show, follow \cite{AKV},  that this term  plays
in fact the crucial role for the reducing  decoherence.
Recently Viola and Lloyd \cite{VL}  have proposed
to use the spin-flip transitions for the dynamical suppression
of decoherence. The control procedure in their scheme
is implemented as a sequence of radiofrequency pulses that repetitively
flip the state of the system.

 Turing machines and circuits are two well known models
 of classical and quantum computers \cite{Deu}.
 Models for the hardware of computers
 are given by circuits that compute Boolean functions.
  For a given sequence of Boolean functions $f_n$
one looks for circuits $S_n$ computing $f_n$
with small size and small depth.
 A hardware problem is a Boolean function $f_n$
 and obviously each $f_n$ is computable.
 A software problem is a language $L$, i.e. a subset
 of the set of all finite $0-1$-sequences and it is known that
 many languages are not computable.
 Circuits $S_n$ for $f_n$ are efficient (or uniform)
 if there is an efficient algoriphm for $L$, the union
 of all $f_n^{-1}(1)$.
 A circuit works only
 for inputs of a definite length whereas a reasonable
 program works for inputs of arbitrary length.
Models for the software of computers are given by Turing machines.
A universal TM models programmable computers.
Uniform circuits can be simulated efficiently by Turing machines.

 In the next section we discuss mathematical models
 of quantum computers such as a multidimensional Turing machine
 and quantum circuits and its relations with spin and gauge theories
  on the lattice. Then the problem of decoherence is considered.

\section{Quantum Turing machine }

A quantum Turing machine (QTM) is obtained by quantization
of a classical Turing machine (TM), so let us first
describe the classical TM. If $\Sigma$ is an alphabet, i.e.
a finite set of simbols, then $\Sigma^*$ denotes the set of
all strings (words) over $\Sigma$. Let $\Sigma$ and $Q$
be two different alphabets.

{\it A Turing machine} $M$
is a triple $M=\{Q,\Sigma,\delta\}$. Here $Q$ is a set of {\it states},
including the initial state $q_0$ and the final state $q_1$
and $\Sigma$ is the {\it tape alphabet} including a blank symbol $a_0$.
The program is given by the transition function

$$
\delta :(Q-q_1)\times \Sigma \to Q\times\Sigma\times\{R,L,N\}
$$
The tape of TM is a sequence of cells, i.e. the one-dimensional
lattice {\bf Z}. Each cell of the tape contains a letter
of $\Sigma$. At the beginning the input $x=(x_1,...,x_n)$
is contained in the cells $0,...,n-1$, all other cells
contain the blank symbol $a_0$.
The central unit of the machine is at each point
of time in one state $q\in Q$. The state at the beginning of computation
if defined by $q_0$. The machine has a head which can read one cell
and which can move from the cell $i\in {\bf Z}$ to the cell $i+1$
or $i-1$. The head starts in the state $q_0$.
If the machine is in state $q$ and reads $a$ and if
$\delta (q,a)=(q',a',\sigma)$, then the machine replaces
the contents of the considered cell by $a'$, the new state of machine
is $q'$, and the head moves one step in direction $\sigma$ (R= right,
L= left, N= no move). The computation stops in the state $q_1$.
The result of computation can be read consecutively in the cells
starting at $0$ until the first cell contains the letter $a_0$.

For input $x$ let $t(x)$ be the number of steps untill the machine
stops and let $s(x)$ be the number of different cells which
are scanned by the head of the machine.
The computational time $t$ and the space $s$ of the Turing machine
are defined by
$$
t(n)=max\{t(x)| |x|=n\}
$$
where $|x|$ is the length of $x$ and
$$
s(n)=max\{s(x)| |x|=n\}
$$
A TM decides a language $L$
if it computes $1$ if $x\in L$ and $0$ if $x\notin L$.
A configuration of TM is a vector
$$
\label{4}
K=(i,q,a_{j_1},...a_{j_p})
$$
where $i\in {\bf Z}$ is the position of the head on the tape,
$q\in Q$ is the current state and $a_j\in \Sigma$ is the contents
of the cell $j$. Equivalently a configuration can be written as
\be
\label{A}
K=(i,q,F)
\ee
where $F:{\bf Z}\to\Sigma$ is a function with only a finite set
of values different from $a_0$.
A multidimensional tape TM has a $d$-dimensional tape ${\bf Z}^d$.
A $d$-dimensional tape TM can move its head in $2d$ directions.

Now let us perform  quantization of the classical Turing machine.
Let $\Sigma$ and $Q$ be two languages described above.
The Hilbert space ${\cal H}$ of states of quantum Turing machine (QTM)
is formed by complex vectors $\psi=\{\psi_K\}, \psi_K\in {\bf C}$
indexed by configurations $K$ of the classical TM with the scalar product
$$
(\psi,\phi)=\sum_K {\bar \psi}_K\phi_K
$$
In ${\cal H}$ there is an orthonormal basis $\{e(K)\},~(e(K),
e(P))=\delta_{KP}$ where $e(K)_{K'}=\delta_{K'}^K$. Because of (\ref{A})
one can write $e(K)=e(i,q,F)$.

{\it A quantum Turing machine } $M$ is a 4-tuple
$M=\{Q,\Sigma,{\cal H},U\}$ where ${\cal H}$ is the Hilbert space
described above and $U$ is a unitary operator in ${\cal H}$ that
satisfies the following condition. There exist three
functions
$$
u_{\sigma}:Q\times\Sigma\times Q\times\Sigma\to{\tilde {\bf C}}
,~~\sigma=0,\pm 1
$$
such that one has the following locality condition
\be
\label{2}
(e(i,q,F),Ue(i',q',F'))=[\delta_{i'}^{i+1}u_1(z)+
\delta_{i'}^{i-1}u_{-1}(z)+\delta_{i'}^i u_0(z)]\prod_{j\neq i}
\delta_{F(i)}^{F'(j)}
\ee
for all configurations $(i,q,F)$ and $(i',q',F')$. Here $z=(q,F(i),q',F'(i))$.
The product
     $\prod_{j\neq i}\delta_{F(j)}^{F'(i)}$
     nonvanishes only in the case if
functions $F$ and $F'$ are
perhaps different at the point $i$. Here ${\tilde {\bf C}}$
is the field of complex numbers computable in the polynomial time.
One can prove that it is enough to restrict ourself
to the field of rational numbers. The fundamental role of
the rational numbers is emphasized also in the p-adic
approach to mathematical physics \cite{VVZ}. It would be very
interesting to explore interrelations between these two approaches.

Unitary evolution operator $U_t$ is defined as the
product of $U$, $U_t=U...U,
t=0,1,...$.  Computations on QTM are performed as follows. Let
$i=0$ and $q_0$ corresponds to the input
$F_0$ and let $e(0,q_0,F_0)$ be the initial vector.
We denote $\psi_t=U_te(0,q_0,F_0)$. If the vector $\psi_t$
admits an  expansion of the form
$$
\psi_t=\sum_{F}c(F)e(0,q_1,F)
$$
where every  vector $e(0,q_1,F)$ from the expansion includes
the final state $q_1$, and moreover the vector $\psi_{\tau}$ does not
admit  such an expansion for any $\tau<t$,  then the computation
is considered the finished in time $t$ and the vector $\psi_t$
is the result of computation.  In this case the probability
to obtain the result
 $F$  is $|c(F)|^2/N$ where $N=\sum |c(F)|^2$.

The classical TM can be considered as a particular case
of QTM when the functions $u_{\sigma}$ take values 0 and 1.

We have discussed the  QTM whose tape
is  the one-dimensional
lattice ${\bf Z}$.
The crucial property of the unitary evolution operator $U$
is the locality condition (\ref{2}).
A QTM provides perhaps not the best model for quantum computations.
One could argue that models of quantum
computations with a local Hamiltonian and not with the local
evolution operator would be  more
realistic physically.

 To extend the efficiency of
quantum computations we can look for a QTM with the multidimensional
tape. It is known that a classical TM with the multidimensional tape
is polynomially equivalent to a TM with the one-dimensional tape.
From the other side properties of one-dimensional and
multidimensional quantum spin systems are drastically different.
Therefore it is worth to study a multidimensional QTM.
We define a QTM $M$
with the multidimensional tape
by quantizing a TM with the multidimensional tape.

{\bf Definition}. {\it A quantum Turing machine $M$
with the multidimensional
tape ${\bf Z}^d$} is a 4-tuple $M=\{Q,\Sigma,{\cal H},U\}$
where the Hilbert space ${\cal H}$ has the basis $\{e(i,q,F)\}$
with $i\in {\bf Z}^d,~q\in Q$ and $F:{\bf Z}^d\to \Sigma$.
  The locality condition reads
$$
(e(i,q,F),Ue(i',q',F'))=[\prod_{\sigma:|i-i'+\sigma|\leq 1}
\delta^{i'}_{i+\sigma}u_{\sigma}(z)]\prod_{j\neq i}
\delta_{F(i)}^{F'(j)}
$$
for all configurations $(i,q,F)$ and $(i',q',F')$. Here $z=(q,F(i),q',F'(i))$.

Quantum Turing machines with two- and three-dimensional tapes
have, perhaps, a better chance for a practical realization
of quantum computer
than the one-dimensional models.

Concerning the complexity of the Ising model it is
known \cite{Wel} that finding a groundstate in the antiferromagnetic
case is $P$ - hard but in the ferromagnetic case the problem is
$NP$ - hard for general graphs and is in $P$ for planar graphs.

 Evolution operators  with certain locality
 conditions on the plane lattice are studied
in the theory of quantum integrable models \cite{FV}.

One has the following example of the QTM.
Let ${\cal H}$ be the Hilbert space with the orthonormal basis
 $e(i,\alpha)$ where
$i\in {\bf Z}$ and $\alpha = 1,2,...,n.$ Let
$U$ be an operator in ${\cal H}$  satisfying
the conditions
$$
(e(i,\alpha),Ue(i',\alpha'))=\delta_{i+1,i'}A_{\alpha\alpha'}
+\delta_{i-1,i'}B_{\alpha\alpha'}
$$
for all $i,i',\alpha,\alpha'$.  Here
$A=(A_{\alpha\alpha'})$
and $B=(B_{\alpha\alpha'})$  are some $n\times n$ matrices.
Then the operator $U$ is unitary if and only if
the matricies $A$ and $B$ satisfy the following conditions:

$$
AA^*+BB^*=I,~
A^*A+B^*B=I,~AB^*=0
$$

A quantum computation is a sequence of unitary transformations.
Any unitary matrix can be approximated by means of the product
of unitary matrices of a simple form. Such a representation is the
quantum analogue of the representation of a recursive function
in terms of primitive functions.
It is known that the set $\{e^{2\pi in\theta}|n\in {\bf Z}\}$,
where $\theta$ is a fixed irrational number, is dense on the unite
circle.
This can be interpreted as saying that
the $1\times 1$ matrix $e^{2\pi i\theta} $ is universal
for the set of all unitary
 $1\times 1$ matrices.

A unitary  $d\times d$-matrix $U$ is of  a {\it simple form},
if it has (after possible reodering)
a block-diagonal form such that every block is a
$2\times 2$-matrix of rotations
$$
\left(
\begin{array} {lr}
\cos \theta & -\sin \theta \\
\sin \theta & \cos \theta
\end{array}  \right)
$$
or it is a number $e^{i\theta}$ with some $\theta$.

One denotes $QMT_{\theta}$ the subset of QTM
with the evolution operator of the described form.

  There exists a classical algoriphm \cite{BV}
(i.e. a classical TM) that for a given $d\times d$-matrix
$U$  and $\epsilon >0$ it computes in the time polynomial
in $d$ and $\log 1/\epsilon$ unitary matrices
of the simple form  $U_1,...,U_n$, where  $n$ depends
polynomially on $d$, such that
$$
||U-U_1\cdots U_n||<\epsilon
$$
Therefore to perform an arbitrary quantum computation
one has to build  a QTM performing the simple unitary
transformations.
There exists a {\it universal} QTM ${\cal M}$
that performs an approximate computation of simple
unitary transformations with an arbitrary small error.
Moreover the unitary evolution operator of QTM
${\cal M}$  includes only the block-matrices
of rotations on the fixed angle. The Hilbert space ${\bf C}^2$
is called {\it qubit}. Any $d$-dimensional
unitary matrix may be written as a product of $2d^2-d$
unitary matrices ({\it quantum gates}),
 each of which acts only within a two-qubit
subspace.

Let be given two QTM $M_1$ and $M_2$ with the same
tape alphabet $\Sigma$. One says that QTM $M_2$
approximates QTM $M_1$
if the following condition is satisfied.
Let $M_1$ for an input $e(0,q_0,F)$ computes the output $\psi_1$
in time $t$. Then $M_2$  for the same input
computes the output $\psi_2$ in time polynomial in $t$ and moreover
$$
||\psi_1 -\psi_2||<\epsilon.
$$

A universal QTM should approximate any QTM. To this end
one has to enumerate and encode
all QTM in such a way that any QTM can be considered
as an input for the universal QTM. Every QTM
can be represented as a finite set $\{Q,\Sigma,u_{\sigma}(q,a,q',a')\}$.
All such  sets  can be enumerated by using the tape
alphabet $\{0,1,a_0\}$.

   There exists a QTM ${\cal M}$ (universal QTM) \cite{BV}
that approximates any QTM  .
Moreover there exists a universal QTM ${\cal M}$ belonging
 to the class $QTM_{\theta}$,
where the number  $\theta$, $0<\theta<\pi/2$ is such that
$\cos\theta$ and  $\sin\theta$ are rational numbers;
in particular one can take  $\cos\theta=3/5,~\sin\theta=4/5$.

Therefore for the practical realisation of quantum computers
it is sufficient to implement rotations on two qubits subsystems
of a quantum system.

\section {Quantum circuits}

Quantum circuits are quantum analogues of the classical circuits
computing Boolean functions. A classical circuit can be represented
as a directed acyclic graph. Similarly a quantum circuit is a sequence
of unitary matrices of the special form associated with a (hyper)graph.

Let us consider a quantum system with the Hilbert space
${\cal H}$ from $l$ qubits, i.e. ${\cal H}=H^{\otimes l}$,
where qubit $H={\bf C}^2$.
It is convenient to treat ${\cal H}$ as the set of functions
on the set
 $L=\{1,2,...,l\}$
with values in $H$, i.e.
${\cal H}=H^{\otimes l}=H^L$. If $g$ is a subset of
$L,~g\subset L$, then one has
$H^L=H^g\otimes H^{L\backslash g}$. Let $U$ be a unitary operator in
$H^{\otimes r},~r\leq l$ and the cardinality of the set  $g\subset L$
is $r=|g|$. Let us denote  $U^{(g)}$
the corresponding unitary operator in $H^g$ and let $\Lambda_g(U) $
be its extension to a unitary operator in $H^L$ obtained by tensoring
to the identity, i.e.
$\Lambda_g(U)=U^{(g)}\otimes id|(H^{L\backslash g}) $.
Let ${\cal U}=\{V,W,...\}$ be a finite set  of unitary
operators (basis of quantum gates) each of which acts in one of the spaces
 $H^{r_{\alpha}},~r_{\alpha}\leq l,~\alpha=1,...,q$.

A pair $(L,\Gamma)$ where $L$ is a set
and $\Gamma$ is a family of its subsets
is called {\it a hypergraph}.
If one of the members of the family is marked then
such a hypergraph is called a network or circuit.
We will deal with the hypergraph
 $(L,\Gamma)$,
where $L=\{1,...,l\}$ and $\Gamma=\{L_1,...,L_T\}$ is a family
of subsets from ${L,~L_i\subset L}$.

{\bf Definition}. {\it  A quantum curcuit} $S_l$ is a 5-tuple
$$
S_l=\{(L,\Gamma),~H^L,~{\cal U},~U_{\Gamma}\}
$$
Here $(L,\Gamma),~H^L$ and ${\cal U}$ are the described above hypergraph,
Hilbert space and the basis of gates,  and
$U_{\Gamma}$  is a unitary operator in  $H^L$ of the form
$$
U_{\Gamma} = \Lambda_{L_T}(U_T)...\Lambda_{L_1}(U_1),
$$
where $U_i$ are unitary operators from the basis ${\cal U}$.
It is assumed that the hypergraph and the basis are consistent in the sence
that  $U_i$ acts in
 $H^{r_{\alpha(i)}}$ and $|L_i|=r_{\alpha(i)},~i=1,...,T$.
Here $\alpha$ is a  function from $\{1,...,T\}$
to $\{1,...,q\}$.

If $\{e_a\}$,~$ a=0,1$ is a basis in qubit ${\bf C}^2$
then the unitary operation in ${\bf C}^4$ of the form
$e_a\otimes e_b\to e_a
\otimes e_{a\oplus b}$ is called the quantum controlled NOT gate.
Here $\oplus$ denotes addition modulo 2.  One can show
that the quantum controlled NOT gate, together with one-qubit gates,
is sufficient for any arbitrary quantum computation.

By using the Dirac notations one writes $|a_1,...,a_l>=e_{a_1}
\otimes ...\otimes e_{a_l}$. The basis $\{|a>=|a_1,...,a_l>\}$ in $H^l$
is called the computational basis. Here $a=\sum_{i=o}^{l-1}2^ia_i$
is the binary decomposition on the number $a$.

More details about mathematical models of quantum computers
one can find for example in \cite{Vol2}.

\section{Reducing of decoherence in quantum computers}

The maintenance of quantum coherence is a crucial
requirement of the ability of quantum computers to be more
efficient in certain problems than classical computers.
One can simulate the environment as a classical
or quantum white noise \cite{HKP}. In a simple model
of spin coupling with the massless quantum field
it was found \cite{Unr} that for quantum computations
not only the coupling with the environment must be small,
but the time taken in the quantum calculation must be less
than the thermal time scale $1/ T$ where $T$ is the temperature
of the initial state of the environment.

The tape (memory) cells are taken to be two-level systems (qubits),
 with each of the levels having the same energy
and  the two states
 are taken to be the eigenstates of the spin operator $\sigma_z$.

Any (pure or mixed) state of a quantum circuit $S_l$
can be described
by a density operator of the form
\be
\rho_S (t)=\sum_{a,b=0}^{2^l-1} \rho_{ab}(t)|a><b|
\ee
where $\{|a>\}$ is the computational basis in $H^l$.
The degree of the quantum coherence is described
by the off-diagonal elements $\rho_{ab},a\neq b$ of the density operator.
The decoherence is the decay of the off-diagonal elements
of the density operator in the computational basis.

In the simple model of the quantum computer interacting
with the environment (reservoir) represented by a quantum field
(a family of harmonic
oscillators) one assumes that the total Hilbert space
is $H^l\otimes {\cal F}$ where $H^l$ is the $l$-qubit space while
${\cal F}$ is the bosonic Fock space. If $\rho_{Tot}(t)$ is the density
operator of the total system then to get the density
operator  $\rho_S(t)$ of the quantum computer one has to take
the partial trace over the reservoir space
$$
\rho_S(t)=Tr_{{\cal F}}(\rho_{Tot}(t))
$$

The Hamiltonian decsribing the coupling of qubit with the
environment
(the spin--boson Hamiltonian) has the form
\be
\label{5}
 H_\lambda=-{1\over2}\Delta\sigma_x+{1\over2}\,\epsilon\sigma_z+\int
dk\omega(k)a^+(k)a(k)+\lambda\sigma_z(A(g^*)+A^+(g))
\ee
where $\sigma_x$ and $\sigma_z$ are Pauli matrices, $\epsilon$
and $\Delta$ are
real parameters interpreted respectively as the
energy  of the spin and the spin-flip parameter. Here
$$A^+(g)=\int a^+(k)g(k)dk,~~A(g^*)=\int a(k)g^*(k)dk$$
where $a(k)$, and $a^+(k)$ are bosonic annihilation and creation
operators
$$[a(k),a^+(k')]=\delta(k-k')$$
which describe the environment.

The one--particle energy of the environment is denoted
$\omega(k)$  and we assume
$\omega(k)\geq 0$. The function $g(k)$ is a form factor describing the
interaction of the
system with the environment, $\lambda$ is the coupling constant.
It is well known that, in times of order
$t/\lambda^2$, the interaction produces effects of order $t$. Thus
$\lambda$ provides a natural time scale for the observable effects of
the interaction system--environment.

Leggett et al. \cite{LCG} have found a very rich behavior
of the dynamics of the
Hamiltonian (\ref{5})
ranging from undamped oscillations, to exponential relaxation, to
power--law types of behavior and to total localization.
Main qualitative features of the
system dynamics can be described in terms of the
temperature (i.e. the initial state of the environment) and of the
behavior, for low frequencies $\omega$, of the spectral function
\be
\label{6}
J(\omega)=\int dk|g(k)|^2\delta(\omega(k)-\omega)
\ee
The dynamics of the Hamiltonian
(\ref{5}) in the so called {\it stochastic approximation}
have been considered in \cite{AKV}.It was found
that the pure oscillating regime when  no damping
and no decoherence is present is descibed by the simple equation
\be
\label{7}
J(\nu\Delta)=0
\ee
where
$$
\nu=\sqrt{1+\left({\epsilon
\over\Delta}\right)^2}
$$

The basic idea of the stochastic approximation (see \cite{ALV})
is the following.
If one has a Hamiltonian of the form
\be
\label{8}
H_\lambda=H_0+\lambda V
\ee
then the stochastic limit of the evolution operator
$$U^{(\lambda)}(t)=e^{itH_0}e^{-itH_\lambda}$$
is the following limit (when it exists in the sense of the convergence of
matrix elements):
$$U(t)=\lim_{\lambda\to\infty}U^{(\lambda)}\left({t\over\lambda^2}\right)
$$
The
limiting evolution operator $U(t)$  describes the
behavior of the model in the time scale $t/\lambda^2$.

In order to apply the stochastic approximation to the Hamiltonian (\ref{5}),
we write (\ref{5}) in the form (\ref{8}) where
$$H_0=H_S+H_R$$
The system Hamiltonian $H_S$ is
$$H_S=-{1\over2}\Delta\sigma_x+{1\over2}\epsilon\sigma_z$$
and the reservoir Hamiltonian $H_R$ is
$$H_R=\int dk\omega(k)a^+(k)a(k)$$
The evolution operator $U^{(\lambda)}(t)$ satisfies equation
$${dU^{(
\lambda)}(t)\over dt}=-i\lambda V(t)U^{(\lambda)}(t)$$
where $V(t)=e^{itH_0}V e^{-itH_0}$ has the form
$$V(t)=\sigma_z(t)(A(e^{-it\omega}g^*)+A^+(e^{it\omega}g))$$
and
\be
\label{9}
\sigma_z(t)=e^{itH_S}\sigma_ze^{-itH_S}
\ee
Let us compute (\ref{9}). The eigenvalues
of the Hamiltonian $H_S$ are
$$H_S|e_\pm>=\lambda_\pm|e_\pm>$$
where
$$\lambda_\pm=\pm{1\over2}\,\Delta\nu,~~
|e_\pm>={1\over\sqrt{1+\mu^2_\mp}}\pmatrix{
1\cr
\mu_\mp\cr}$$
and
$$\mu_\pm={\epsilon\over\Delta}\,\pm \nu\ ,\quad \nu=\sqrt{1+\left
({\epsilon
\over\Delta}\right)^2}$$
Notice  that:
$$\langle e_\pm|\sigma_z|e_\pm\rangle={1-\mu^2_\mp\over1+\mu^2_\mp}\quad;
\qquad
\langle e_+|\sigma_z|e_-\rangle=\langle e_-|\sigma_z|e_+\rangle=1/\nu$$
Therefore
$$\sigma_z(t)={1-\mu^2_-\over1+\mu^2_-}\,DD^++{1-\mu^2_+\over1+\mu^2_+}\,
D^+D+\nu^{-1}e^{it \nu\Delta}\,D+ \nu^{-1}e^{-it\nu\Delta} D^+
$$
where
$$D=|e_+><e_-|$$
The interaction Hamiltonian  can now be written in the form :
$$V(t)=\sum^3_{\alpha=1}(D^+_\alpha\otimes A(e^{-it\omega_\alpha}
g^*)+h.c.)$$
where the three spectral frequencies correspond respectively to the
down, zero, and up transitions of the 2--level system, i.e.
$$\omega_1(k)=\omega(k)-\nu\Delta\quad ,\qquad
\omega_2(k)=\omega(k)\quad ,\qquad\omega_3(k)
=\omega(k)+\nu\Delta$$
$$D_1=\nu^{-1}D^+\quad ,
\qquad D_2={1-\mu^2_-\over1+\mu^2_-}\ DD^++{1-\mu^2_+\over1+
\mu^2_+}\, D^+D\quad ,\qquad D_3=\nu^{-1}D^+$$
The spectral frequencies $\omega_2(k)$ and $\omega_3 (k)$
are positive. Therefore in the stochastic limit
one gets only one white noise field. Still
a remnant of the interaction remains because, after the limit, the system
evolves with a new hamiltonian, equal to the old one plus a shift term
depending on the interaction and on the initial state of the field. This
was called in \cite {AKV} a {\it Cheshire Cat effect}.

 The limiting evolution equation can then be written:
\be
\label{10a}
{dU(t)\over dt}\,=Db^+(t)U(t)-D^+U(t)b(t)-(\gamma+i\sigma)
D^+DU(t)-i\varphi U(t)
\ee
where
$$\gamma=\nu^{-2}\pi J(\nu\Delta)\ ,$$
$$\sigma=\nu^{-2}(I(-\nu\Delta)-I(\nu\Delta))+\left(\left({1-\mu^2_-\over1+
\mu^2_-}\right)^2-\left({1-\mu^2_+\over1+\mu^2_+}\right)^2\right)I(0)\ ,$$
$$\varphi=\nu^{-2}I(-\nu\Delta)+\left({1-\mu^2_-\over1+\mu^2_-}
\right)^2I(0)$$
and we denote
$$J(\omega)=\int dk|g(k)|^2\delta(\omega(k)-\omega)\quad;\qquad
I(\omega)=P.P.\int^\infty_0{d\omega' J(\omega')\over\omega'-\omega}
$$
where $P.P.$ means the principal part of the integral.
The operators
$b(t)$, $b^+(t)$ satisfy the quantum white noise relations
$$[b(t),b^+(t')]=\gamma\delta(t-t')$$

In the notations of quantum stochastic equations the equation (\ref{10a})
reads $$dU(t)=(DdB^+_t-D^+dB_t-(\gamma+i\sigma)D^+D-i\varphi)U(t)$$ Notice
that all parameters $\gamma$, $\sigma$ and $\varphi$ in the evolution
equation (\ref{12}) are expressed in terms of the spectral density
$J(\omega)$ and parameters $\Delta$ and $\epsilon$ of the
original Hamiltonian.

For zero temperature
the stochastic approximation to the vacuum expectation value of the
Heisenberg evolution of $\sigma_z$ is given by
$$P(t)=\langle U^*(t)\sigma_z(t)U(t)\rangle$$
From equation (\ref{10a}) one gets the Langevin equation for $P(t)$ which
solution is
$$
P(t)=\nu^{-1}e^{-\gamma t}(D^+e^{i(\sigma-\nu\Delta)t}+De^{-i
(\sigma-\nu\Delta)t})+
$$
\be
\label{11}
+D^+D\left({1-\mu^2_+\over1+\mu^2_+}\,-{1-\mu^2_-\over1+\mu^2_-}\right)
e^{-2\gamma t}+{1-\mu^2_-\over1+\mu^2_-}
\ee
We obtain the pure oscillating behaviour and no decoherence if
\be
\label{12}
\gamma=\nu^{-2}\pi J(\nu\Delta)=0
\ee

For a non--zero temperature we get a stochastic evolution equation of
the same form as before  only with new constants $\gamma$, $\sigma$
and $\varphi$. More precisely:
$$\gamma=\nu^{-2}\pi J(\nu\Delta)\coth{\beta\nu\Delta\over2}~,$$
$$\sigma=\left[\left({1-\mu^2_+\over1+\mu^2_+}\right)^2-
\left({1-\mu^2_-\over1+\mu^2_-}\right)^2\right](I_+(0)+I_-(0))+$$
$$+\nu^{-2}(I_+(-\nu\Delta)-I_+(\nu\Delta)+I_-(-\nu\Delta)-I_-
(\nu\Delta))$$
where spectral densities are
$$J_+(\omega)={J(\omega)\over1-e^{-\beta\omega}}\qquad;\qquad
J_-(\omega)={J(\omega)e^{-\beta\omega}\over1-e^{-\beta\omega}}$$
Here $J(\omega)$ is the spectral density (\ref{6})
and $\beta$ is the inverse temperature.
The functions $I_\pm(\omega)$ are defined by
$$I_\pm(\omega)= P.P.\int{d\omega'J_\pm(\omega')\over\omega'-
\omega}$$
One has the same as for the zero--temperature expression
(\ref{11}) for
$P(t)$ but  now with new constants $\gamma$ and $\sigma$
depending on temperature.
The condition for the reducing of coherence is still the same (\ref{12}).
It seems this condition is a rather week requirement
on the parameters of the interaction between
quantum computer and the environment, so one can hope to use  it
to reduce decoherence.

\section{Acknowledgments}
I am very grateful to Prof. T. Hida for the invitation
to the Conference on Quantum Information held at Meijo University,
Nagoya, 4-8 Nov. 1997,
with the very exciting scientific environment.


\begin{thebibliography}{99}

\bibitem{Ben} P. Benioff, {\it The computer as a physical system: a microscopic
quantum mechanical Hamiltonian model of computers as represented by Turing
machines}, J.Statist.Phys. 22 (1980) 563-591

\bibitem{Fey} R. Feynman, {\it Quantum mechanical computers},
Found. Phys. 16 (1986) 11-20

\bibitem{Deu} D. Deutsch, {\it Quantum theory, the Church-Turing
principle and the universal quantum computer}, Proc.Roy.Soc.London Ser.A,
400 (1985) 96-117


\bibitem{BV} E. Bernstein and U. Vazirani, {\it Quantum complexity theory},
in Proc. of the 25th Annual ACM Symposium on Theory of Computing, ACM,
New York, 1993, pp.11-20.

\bibitem{Sho} P. Shor, {\it Algorithms for quantum computations: Discrete
logarithms and factoring}, in Proc. of the 35th Annual Symposium on
Foundations of Computer Science, IEEE Computer Society Press, Los
Alamitos, CA, 1994, pp.124-134

\bibitem{BBC} A. Barenco, C.H. Bennett, C. Cleve, D.P. DiVincenzo,
N. Margolius, P. Shor, T. Sleator, J.A. Smolin and H. Weinfurter,
{\it Elementary gates for quantum computations}, Phys. Rev.A, 52 (1995)
3457-3467

\bibitem{Acc} L. Accardi, {\it An open system approach} {\it to quantum
computers}, in: Quantum Communication and Measurement, O. Hirota,
A. S. Holevo, C. M. Caves (eds.), Plenum Press, 1997, pp.387-393

\bibitem{OW} M. Ohya and N. Watanabe, {\it On mathematical treatment
of Fredkin-Toffoli-Milburn gate}, Physica D,120 (1998) 206-213

\bibitem{Oza} M. Ozawa, {\it Measurability and computability},
LANL e-print quant-ph/9809048

\bibitem{Unr} W. Unruh, {\it Maintaining coherence in quantum computers},
Phys. Rev.A, 51 (1995) 992-997

\bibitem{PSE} G.M. Palma, K.-A. Suominen and A.K. Ekert, Proc.Roy.Soc.
London Ser.A, 452 (1996) 567-574

\bibitem{DS} D.P. DiVincenzo and P.W. Shor, Phys.Rev.Lett. 77 (1996) 3260-3265

\bibitem{ZR} P. Zanardi and M. Rasetti,  Phys. Rev. Lett. 79 (1997) 3306-3311

\bibitem{DG} L.M. Duan and G.C. Guo, Phys. Rev. Lett. 79 (1997) 1953-1958

\bibitem{Vol1} I.V. Volovich, {\it Quantum Computers and Neural Networks},
Invited talk at the  International  Conference on
Quantum Information held at Meijo University,
4-8 Nov. 1997

\bibitem{AKV} L. Accardi, S.V. Kozyrev and I.V. Volovich,
{\it Dynamics of dissipative two-state systems in the stochastic
approximation}, Phys.Rev.A, 56 (1997) N3

\bibitem{VL} L. Viola and S. Lloyd, {\it Dynamical suppression
of decoherence in two-state quantum systems}, LANL e-print quant-ph/9803057

\bibitem{VVZ} V.S. Vladimirov, I.V. Volovich and E.I. Zelenov,
{\it p-Adic Analysis and Mathematical Physics }, World Scientific, 1996

\bibitem{Wel} D.J.A. Welsh, {\it Complexity: Knots, Colouring and Counting},
Cambridge University Press, 1993

\bibitem{FV}  L.D. Faddeev and  A.Yu. Volkov, {\it Algebraic quantization
of integrable models in discrete space-time}, LANL e-print hep-th/9710039

\bibitem{Vol2} I.V. Volovich, {\it Mathematical Models of Quantum Computers},
Lectures delivered at the Moscow State University, 1998, to be published

\bibitem{HKP} T. Hida, H.-H. Kuo, J. Potthoff and L. Streit,
{\it White Noise. An Infinite Dimensional Calculus},
Kluwer Academic Publishers,
(1993)

\bibitem{LCG} A.J. Leggett, S. Chakravarty, A. Garg and L.D. Chang,
Rev.Mod.Phys.59 (1987) N 1


\bibitem{ALV}  L. Accardi, Y.G. Lu and I.V. Volovich, {\it Quantum
Theory and Its Stochastic Limit}, 1999, Oxford University
Press (to be published)


\end{thebibliography}
\end{document}